\label{key}%\documentclass[11pt]{article}

\documentclass[preprint, showpacs, showkeys]{revtex4-1}
\usepackage{titlesec}
\usepackage[T1]{fontenc}

\usepackage{lipsum} %paquete para llenar texto
\usepackage{footnote}
\usepackage{amsmath}
\usepackage{subfigure} % paquete para colocar 2 imagenes en una sola
\usepackage[ansinew]{inputenc}
\usepackage{float}
\usepackage{graphicx}% Include figure files
\usepackage{dcolumn}% Align table columns on decimal point
\usepackage{bm}% bold math
\usepackage{amssymb}
\usepackage{verbatim}
\usepackage{multirow}
\usepackage{booktabs}
\usepackage{soul}
\usepackage{color}
\definecolor{Blue}{rgb}{0,0,.7}
\definecolor{Red}{rgb}{0.72,0,0}
\definecolor{Green}{rgb}{0,.395,0}
\definecolor{Pink}{RGB}{255,0,128}
\definecolor{Orange}{RGB}{255,128,0}
\definecolor{LBlue}{RGB}{10,113,179}
\usepackage[colorlinks=false]{hyperref}

\renewcommand\thesection{\arabic{section}}
\renewcommand\thesubsection{\thesection.\arabic{subsection}}
\usepackage[none]{hyphenat}
\newcommand{\vvector}[1]{\mathbf{#1}}
\newcommand{\B}{\mathbf{B}}
\newcommand{\operador}[1]{\hat{#1}}
\newcommand{\oH}[1]{\operador{H}_{#1}}
\newcommand{\x}{\operador{x}}
\newcommand{\y}{\operador{y}}
\newcommand{\z}{\operador{z}}
\newcommand{\p}{\operador{p}}
\newcommand{\px}{\operador{p_x}}
\newcommand{\py}{\operador{p_y}}
\newcommand{\pz}{\operador{p_z}}
\newcommand{\A}{\vvector{A}(\operador{r})}
\newcommand{\punto}[2]{#1 \cdot #2}
\newcommand{\oaa}{\operador{a}^+}
\newcommand{\oa}{\operador{a}}
\newcommand{\obb}{\operador{b}^+}
\newcommand{\ob}{\operador{b}}
\newcommand{\oAA}{\operador{A}^+}
\newcommand{\oA}{\operador{A}}
\newcommand{\oBB}{\operador{B}^+}
\newcommand{\oB}{\operador{B}}
\newcommand{\N}[1]{\operador{N}_#1}
\titleformat{\subsection}
{\bfseries \itshape \fontsize{6}{8}}{\thesubsection}{1em}{}

\begin{document}
\setcounter{page}{0}
\title[]{Energy levels in a single-electron quantum dot with hydrostatic pressure}
\author{H. E. Caicedo-Ortiz}\email{Corresponding Author, E-mail: hernando.caicedo@uniautonoma.edu.co}
\affiliation{Corporaci\'on Universitaria Aut\'onoma del Cauca, Popay\'an, Colombia}
\affiliation{Departamento de Ciencias B\'asicas, Facultad de Ingenier\'ia, Universidad Anahuac Norte, M\'exico}
\author{H. O. Casta\~neda Fern\'andez}
\affiliation{Escuela Superior de F\'isica y Matem\'aticas, Instituto Polit\'ecnico Nacional, M\'exico}
\author{E.  Santiago-Cort\'es}
\affiliation{Corporaci\'on Universitaria Aut\'onoma del Cauca, Popay\'an, Colombia}
\affiliation{Escuela Nacional de Ciencias Biol\'ogicas y Centro de Nanociencias y Micro y Nanotecnolog\'ias, \\Instituto Polit\'ecnico Nacional, M\'exico}
\author{D. A.  Mantilla-Sandoval}
\affiliation{Corporaci\'on Universitaria Aut\'onoma del Cauca, Popay\'an, Colombia}

\newpage
\date{\today}

\begin{abstract}
In this article we present a study of the effects of hydrostatic pressure on the energy levels of a quantum dot with an electron.  A quantum dot is modeled using an infinite potential well and a two-dimensional harmonic oscillator and solved through the formalism of second quantization. A scheme for the implementation of a quantum NOT gate controlled with hydrostatic pressure is proposed. 
\end{abstract}

%\pacs{68.65.Hb, 68.55.ag, 03.67.Lx}

\keywords{quantum dots; harmonic oscillator; potential well; quantum gate; quantum computer}

\maketitle

\section{Introduction}

Semiconductor quantum dots (QDs), are low dimensional systems, in which charge
carriers are confined to three dimensions \cite{ekimov,jacak,jacak2,masumoto,tartakovskii,reimann,caicedo2015}, and are composed mainly of GaAs, GaAsAl, CdSe, or
PbS. The quantum dot acts as a box that confines particles like electrons, holes or excitons,
where the number of particles controlled by applying a potential difference across
two electrodes connected to the system. The confinement of the created particles 
produces a discrete quantization of energy levels, which involves changes in the electrical
and optical properties of the system, allowing new possibilities in the design of artificial
atoms and molecules.

Previous work has been done on the effects of hydrostatic pressure on the energy behavior of quantum dots with multiple electrons, with the use of different geometries \cite{ortakaya, duque2016,peter2005,perez2006,perez2008,liang2011,bzour2017,jeice2016,sivakami2010,owji2016,lepkowski2006,bzour2017,caicedo2012interaccionespin}.  Photoluminescence measurement under high hydrostatic pressure has proven to be a useful tool for exploring the electronic structure and optical transitions in quantum dots \cite{tang,manjon}. Zhou et. al.\cite{zhou}  report an experimental study on the optical properties of the self-organized 1.55-nm InAs/InGaAsP/InP quantum dots under hydrostatic pressure up to 9.5 GPa at 10 K. Duque et al. \cite{duque2016}. Segovia and Vinck-Posada \cite{segovia} consider the implications  of hydrostatic pressure and temperature on the defect mode in the band structure of a one-dimensional photonic crystal. Within the framework of effective-mass approximation, the hydrostatic pressure effects on the donor binding energy of a hydrogenic impurity in InAs/GaAs self-assembled quantum dot(QD) are investigated employing a variational method by Xia et al. \cite{xia}.

 The effect of hydrostatic pressure the binding energy of donor impurities in QW structures is calculated by Elabsy \cite{elabsy}. These effects have also been studied by Zhigang et al. \cite{xiao} and Szwacka \cite{szwacka}. Yuan and collaborators  \cite{yuan} studied the binding energy of hydrogenic impurity associated with the ground state and some low-lying states in a GaAs spherical parabolic quantum dot while taking into account hydrostatic pressure and electric field using the configuration–integration
 method. For the construction of quantum gates and spintronics based devices, it is necessary to control the number and properties of the confined electrons in a quantum dots. Loss and DiVincenzo \cite{loss} proposed an implementation of a universal set of one- and two-quantum-bit gates for quantum computation
using the spin states of coupled single-electron quantum dots. Burkard et al.\cite{burkard}  considered a quantum-gate mechanism based on electron spins in coupled semiconductor quantum dots. Caicedo-Ortiz et al.\cite{caicedo2006exchange} proposed a scheme switching system for two laterally coupled quantum dots operating as a quantum gate 2-qubits. A study of the short time dynamics of a charge qubit encoded in two coherently coupled quantum dots connected to external electrodes has also been studied by \L{}uczak and Bu\l{}ka  \cite{luczak}. The study of hydrostatic pressure in semiconductor states is imperative as a phenomenon that can change the properties of any quantum device dramatically.  We propose a  scheme for the implementation of a quantum NOT gate controlled with hydrostatic pressure. 

In this article, we studied the effects of magnetic fields and hydrostatic pressure in the energy of a quantum dot with a single electron. We applied an algebraic method to find an analytical solution to the energy of the system. In Section \ref{model} we present the Hamiltonian of the system; in Section \ref{solution} we develop the method and the solution of the problem. Section \ref{results} shows that the energy of the system as a function of the hydrostatic pressure and a scheme for the implementation of a quantum NOT gate. Finally, we discuss the implications of this work in Section \ref{conclusiones}.

\section{The model}
\label{model}

The three-dimensional Hamiltonian for a paraboloid quantum dot of GaAn, with the effective mass approximation under the effects of hydrostatic pressure $(P)$ in a uniform magnetic field $\vvector{B} = B\hat{k}$, can be written in a separable form as follows:

%%%%%%%%%%%%%%%%%%%%%%%%%%%%%%%%%%%%%%%%%%%%%%%%%%%
\begin{equation}
\label{eqb1}
\operador{H}=\operador{H_0}+\oH{s},
\end{equation}
%%%%%%%%%%%%%%%%%%%%%%%%%%%%%%%%%%%%%%%%%%%%%%%%%%%
where

%%%%%%%%%%%%%%%%%%%%%%%%%%%%%%%%%%%%%%%%%%%%%%%%%%%%%%%
\begin{equation}
\label{eqb2}
\operador{ H_0}=\frac{1}{2 m^{*}_{(P)}}\left[\p+\frac{q_e}{c}\A\right]^2+\frac{m^{*}_{(P)}\omega_0^2}{2}\left(\x^2+\y^2\right)+u(\z).
\end{equation}
%%%%%%%%%%%%%%%%%%%%%%%%%%%%%%%%%%%%%%%%%%%%%%%%%%%%%%%

%%%%%%%%%%%%%%%%%%%%%%%%%%%%%%%%%%%%%%%%%%%%%%%%%%%%%%%%%%%%%%%%%%%%%%%%%%%%%%%%%%%%%%%%%%%%%
\begin{equation}
\label{eqb3}
\oH{s}=-\operador{\mu}_{s}\cdot \B,
\end{equation}
%%%%%%%%%%%%%%%%%%%%%%%%%%%%%%%%%%%%%%%%%%%%%%%%%%%%%%%%%%%%%%%%%%%%%%%%%%%%%%%%%%%%%%%%%%%%%

$\A=\frac{B}{2}(-\y, \x,0)$ is a vector potential, 
$\p=\px+\px+\pz$ represents the total momentun of the system, $m^{*}_{(P)}$ is the effective mass of the electron in the GaAs. $\frac{1}{2}g_s\mu
\punto{\operador{\sigma}}{\vvector{B}}$ is the 
Zeeman coupling,  $g_s$ is the Land\'e factor or gyromagnetic factor, and  $\operador{\sigma}=\operador{\sigma_x} +
\operador{\sigma_y} + \operador{\sigma_z}$ are the Pauli matrices, where $\operador{\mu}_{s}=-\frac{g_s \mu_B \vvector{S}}{\hbar}$, $\mu_B=\frac{q_e \hbar}{2 m^{*}_{(P)}}$ and $\operador{S}=\frac{\hbar}{2}\operador{\sigma}$.

\section{Solution of the Hamiltonian}
\label{solution}

Using  $[\x,\py]=[\y,\px]=0$, with the cyclotron frequency as  $\omega_c=\frac{q_eB}{m_{(P)}^{*}c}$, the projection of the angular momentum in the $z$-axis is
$\operador{L_z}=\x \py - \y \px$  and
$\Omega^2=\omega_0^2+\frac{1}{4}\omega_c^2$, therefore
%%%%%%%%%%%%%%%%%%%%%%%%%%%%%%%%%%%%%%%%%%%%%%%%%%%
\begin{widetext}
\begin{equation}
\label{eqb6}
\operador{H_0}=\frac{\px^2}{2m^{*}_{(P)}}+\frac{\py^2}{2m^{*}_{(P)}}+\frac{\pz^2}{2m^{*}_{(P)}}+\frac{1}{2}\omega_{c}\operador{L_{z}}+
\frac{1}{2}m^{*}_{(P)}\Omega^2\left(\x^2+\y^2\right)+u(\z),
\end{equation}
\end{widetext}
%%%%%%%%%%%%%%%%%%%%%%%%%%%%%%%%%%%%%%%%%%%%%%%%%%%
in a separable form 

%%%%%%%%%%%%%%%%%%%%%%%%%%%%%%%%%%%%%%%%%%%%%%%%%%%
\begin{equation}
	\label{eqb9}
	\operador{H_0}=\operador{H_\bot}+\operador{H_{||}},
\end{equation}
%%%%%%%%%%%%%%%%%%%%%%%%%%%%%%%%%%%%%%%%%%%%%%%%%%%
where
%%%%%%%%%%%%%%%%%%%%%%%%%%%%%%%%%%%%%%%%%%%%%%%%%%%
\begin{equation}
	\label{eqb10}
	\operador{H_\bot}=\frac{\px^2}{2m^{*}_{(P)}}+\frac{\py^2}{2m^{*}_{(P)}}+\frac{1}{2}\omega_{c}\operador{L_{z}}+
	\frac{1}{2}m^{*}_{(P)}\Omega^2\left(\x^2+\y^2\right),
\end{equation}
%%%%%%%%%%%%%%%%%%%%%%%%%%%%%%%%%%%%%%%%%%%%%%%%%%%
and
%%%%%%%%%%%%%%%%%%%%%%%%%%%%%%%%%%%%%%%%%%%%%%%%%%%
\begin{equation}
	\label{eqb11} \operador{H_{||}}=\frac{\pz^2}{2m^{*}_{(P)}}+u(\z),
\end{equation}
%%%%%%%%%%%%%%%%%%%%%%%%%%%%%%%%%%%%%%%%%%%%%%%%%%%
%%%%%%%%%%%%%%%%%%%%%%%%%%%%%%%%%%%%%%%%%%%%%%%%%%%
 $\operador{H_\bot}$ is the Hamiltonian of a bi-dimensional harmonic oscillator and $\operador{H_{||}}$ is Hamiltonian of a infinite square well potential.

\subsection{Hamiltonian of a bi-dimensional harmonic oscillator}

If   
$\beta=\sqrt{\mu\Omega/\hbar}$, and the operators

%%%%%%%%%%%%%%%%%%%%%%%%%%%%%%%%%%%%%%%%%%%%%%%%%%%%%
\begin{equation*}
	\label{eqb13}
	\oaa =\frac{1}{2}\left[\beta \x-\frac{i\px}{\beta\hbar}\right] ; \ \    \obb =\frac{1}{2}\left[\beta \y-\frac{i\py}{\beta\hbar}\right]
\end{equation*}
\begin{equation*}
	\label{eqb14}
	\oa  =\frac{1}{2}\left[\beta x+\frac{i\px}{\beta\hbar}\right] ;\ \    \ob =\frac{1}{2}\left[\beta \y+\frac{i\py}{\beta\hbar}\right],
\end{equation*}
%%%%%%%%%%%%%%%%%%%%%%%%%%%%%%%%%%%%%%%%%%%%%%%%%%%%%

and 

%%%%%%%%%%%%%%%%%%%%%%%%%%%%%%%%%%%%%%%%%%%%%%%%%%%%%
\begin{equation*}
	\label{eqb18} \oBB  =\frac{1}{\sqrt2}\left(\oaa-i\obb\right); \ \  
	\oAA =\frac{1}{\sqrt2}\left(\oaa+i\obb\right)
\end{equation*}
\begin{equation*}
	\label{eqb19}
	\oB =\frac{1}{\sqrt2}\left(\oa+i\ob\right); \ \    \oA
	=\frac{1}{\sqrt2}\left(\oa-i\ob\right),
\end{equation*}
%%%%%%%%%%%%%%%%%%%%%%%%%%%%%%%%%%%%%%%%%%%%%%%%%%%%%

the Hamiltonian $\operador{H_\bot}$  is:

%%%%%%%%%%%%%%%%%%%%%%%%%%%%%%%%%%%%%%%%%%%%%%%%%%%%%
\begin{equation}
\label{eqb21}
\operador{H_{\bot}}=\hbar\Omega\left(\oAA\oA+\oBB\oB+1\right)
+\frac{\omega_c\hbar}{2}\left(\oAA\oA-\oBB\oB\right).
\end{equation}
%%%%%%%%%%%%%%%%%%%%%%%%%%%%%%%%%%%%%%%%%%%%%%%%%%%%%

If $\oAA\oA = \N{A}$ and $\oBB\oB=\N{B}$ with  eigenvalues $n_A$ and $n_B$ where  $n = n_A+n_B$ and
$m = n_A-n_B$, the energy of the system in $xy$ is

%%%%%%%%%%%%%%%%%%%%%%%%%%%%%%%%%%%%%%%%%%%%%%%%%%%%%
\begin{equation}
\label{eqb26}
E_{\bot}=\hbar\Omega\left(n+1\right)+\frac{\hbar \omega_c}{2}
m.
\end{equation}
%%%%%%%%%%%%%%%%%%%%%%%%%%%%%%%%%%%%%%%%%%%%%%%%%%%%%

\vspace{0.5cm}
%%%%%%%%%%%%%%%%%%%%%%%%%%%%%%%%%%%%%%%%%%%%%%%%%%%%%%%%%%%%%%%%%%%%%%%%%%%%%%%%%%%%%%%%%%%%
%%%%%%%%%%%%%%%%%%%%%%%%%%%%%%%%%%%%%%%%%%%%%%%%%%%%%%%%%%%%%%%%%%%%%%%%%%%%%%%%%%%%%%%%%%%%
%%%%%%%%%%%%%%%%%%%%%%%%%%%%%%%%%%%%%%%%%%%%%%%%%%%%%%%%%%%%%%%%%%%%%%%%%%%%%%%%%%%%%%%%%%%%
%%%%%%%%%%%%%%%%%%%%%%%%%%%%%%%%%%%%%%%%%%%%%%%%%%%%%%%%%%%%%%%%%%%%%%%%%%%%%%%%%%%%%%%%%%%%

\subsection{Hamiltonian of a infinite square well potential}
The potential  $u(z)$ for  $\operador{H_{||}}$ is $u(z)=
	\infty \    \text{if} \  z<-\frac{1}{2}\ \text{or} \ z> $ and $ u(z)= 0$  if $ \frac{-1}{2} \  < z < \frac{1}{2}$,  where  the energy is
	%%%%%%%%%%%%%%%%%%%%%%%%%%%%%%%%%%%%%%%%%%%%%%%%%%%%%%%%%%%%%%%
	\begin{equation}
	\label{}
	E_{||}=\frac{n_z^2\pi^2\hbar^2}{2 m^{*}(P) [R(P)]^2}
	\end{equation}
	%%%%%%%%%%%%%%%%%%%%%%%%%%%%%%%%%%%%%%%%%%%%%%%%%%%%%%%%%%%%%%%%
	\subsection{Hamiltonian with spin effect}
	Using $\operador{\mu}_{s}=-\frac{g_s \mu_B \vvector{S}}{\hbar}$, $\mu_B=\frac{q_e \hbar}{2 m_{(P)}^{*}}$ and  $\operador{S}=\frac{\hbar}{2}\operador{\sigma}$, the Hamiltonian \ref{eqb11} is
	
	%%%%%%%%%%%%%%%%%%%%%%%%%%%%%%%%%%%%%%%%%%%%%%%%%%%%%%%%%%%%%%%%%%%%%%%%%%%%%%%%%%%%%%%%%%%%%
	\begin{equation}
	\label{} \oH{s}=\frac{g_s q_e \hbar}{4 m_{(P)}^{*}}\operador{\sigma}\cdot
	\B.
	\end{equation}
	%%%%%%%%%%%%%%%%%%%%%%%%%%%%%%%%%%%%%%%%%%%%%%%%%%%%%%%%%%%%%%%%%%%%%%%%%%%%%%%%%%%%%%%%%%%%%
	If $(g_s=2)$, the energy of the system is
	%%%%%%%%%%%%%%%%%%%%%%%%%%%%%%%%%%%%%%%%%%%%%%%%%%%%%%%%%%%%%%%%%%%%%%%%%%%%%%%%%%%%%%%%%%%%%
	\begin{equation}
		E_{s}=\pm \frac{q_e\hbar}{2 m_{(P)}^{*}} B
		\label{spin01}
	\end{equation}
	%%%%%%%%%%%%%%%%%%%%%%%%%%%%%%%%%%%%%%%%%%%%%%%%%%%%%%%%%%%%%%%%%%%%%%%%%%%%%%%%%%%%%%%%%%%%%

The total energy of the system is 

%%%%%%%%%%%%%%%%%%%%%%%%%%%%%%%%%%%%%%%%%%%%%%%%%%%%%%%%%%%%%%%%%%%%%%%%%%%%%%%%%%%%%%%%%%%%%
\begin{equation}
	E=\hbar\Omega\left(n + 1\right)+\frac{\hbar q_eB}{2m_{(P)}^{*}c}\left(m
	\right) + \frac{n_z^2\pi^2\hbar^2}{2R_{(P)}^2 m_{(P)}^{*}} \pm \frac{q_e\hbar}{2
		m_{(P)}^{*}} B.
	\label{energyeq}
\end{equation}
%%%%%%%%%%%%%%%%%%%%%%%%%%%%%%%%%%%%%%%%%%%%%%%%%%%%%%%%%%%%%%%%%%%%%%%%%%%%%%%%%%%%%%%%%%%%%

\section{Results and discussion}
\label{results}
	
The application of hydrostatic pressure modifies the dot size and effective mass. The variation of the dot with pressure is given by
	
	%%%%%%%%%%%%%%%%%%%%%%%%%%%%%%%%%%%%%%%%%%%%%%%%%%%%%%%%%%%%%%%%%%%
	\begin{equation}
	\label{eqb4}
	R_{(P)}= R_{0}(1-1.5082 \times 10^{-4}P),
	\end{equation}
	%%%%%%%%%%%%%%%%%%%%%%%%%%%%%%%%%%%%%%%%%%%%%%%%%%%%%%%%%%%%%%%%%%%
	
where $P$  is in kbar and $R_{0}$ is the radius value of the quantum dot when the hydrostatic pressure is zero. The effective mass in the  quantum dot is:

	%%%%%%%%%%%%%%%%%%%%%%%%%%%%%%%%%%%%%%%%%%%%%%%%%%%%%%%%%%%%%%%%%%%
	\begin{equation}
	\label{eqb5}
	m_{(P)}^{*}=m_{0}^{*}e^{0.0078P}, 
	\end{equation}
	%%%%%%%%%%%%%%%%%%%%%%%%%%%%%%%%%%%%%%%%%%%%%%%%%%%%%%%%%%%%%%%%%%%
	
	where $R_{0}$ and  $m_{0}^{*}$ are the zero-pressure dot radius and initial effective mass respectively.
	
	%%%%%%%%%%%%%%%%%%%%%%%%%%%%%%%%%%%%%%%%%%%%%%%%%%%%%%%%%%%%%%%%%%%
	%%%%%%%%%%%%%%%%%%%%%%%%%%%%%%%%%%%%%%%%%%%%%%%%%%%%%%%%%%%%%%%%%%%
	%%%%%%%%%%%%%%%%%%%%%%%%%%%%%%%%%%%%%%%%%%%%%%%%%%%%%%%%%%%%%%%%%%%
	
To include the effect of the pressure $(P)$ on the quantum dots energy states and the magnetization, we replaced the effective mass and dot radius as defined by equations \ref{eqb4} and \ref{eqb5} in the quantum dot energy equations \ref{energyeq}.

%\begin{widetext}
\begin{equation*}
\label{h33} 
E=\frac { \hbar \omega _{ 0 }^{*}e^{ -0.0078P } }{ (1-1.5082\times 10^{ -4 }P)^{ 2 } } r  \left( { \left[ 1+\frac { \left( \gamma  \right) ^{ 2 }(1-1.5082\times 10^{ -4 }P)^{ 4 } }{ 4 }  \right]  }^{ 1/2} \left[ n+1 \right]  +\right.
\end{equation*}
\begin{equation}
\label{h33} 
\left. +  \gamma (1-1.5082\times 10^{ -4 }P)^{ 2 }\left[ \frac { m\pm c }{ 2 }  \right] +\frac { { n }_{ z }^{ 2 }{ \pi  }^{ 2 } }{ 2 } \right) 
\end{equation}
%\end{widetext}	

where $\gamma = \frac{\omega_c^{*}}{\omega_0^{*}}$, $\omega_{c}^{*} = \frac{q_{e} B}{m^{*}_{0} c }$ and $\omega_{0}^{*} =\frac{\hbar}{m_{0}^{*} R_{0}^{2}}$. The variable  $\gamma$ is a function of the magnetic field value in the  $z$ direction.

\begin{figure}[h]
		\begin{center}
			\includegraphics[width=8.7cm]{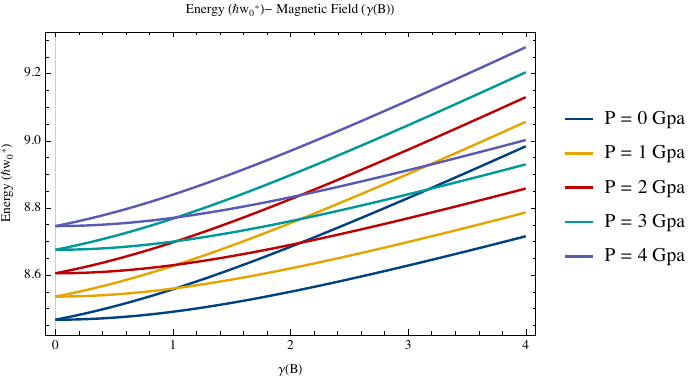}
		\end{center}
		\vspace{-0.5cm}
		\caption{Energy behavior of a quantum dot of constant radius  $\rho$, under the action of an external magnetic field $\B$
			parallel to the $z$ axis, in function of $\omega_c/\omega_0$.}
		\label{1}
	\end{figure}

 In the Figure \ref{1} it is observed how the energy levels present a degeneracy when $\gamma (B)$ increases. This is the result of increasing the magnetic field $\B$ for a pressure $P = 0$. The results are in agreement with the results of Caicedo-Ortiz  \textit{et al.}\cite{caicedo2015}.

		%%%%%%%%%%%%%%%%%%%%%%%%%%%%%%%%%%%%%%%%%%%%%%%%%%%%%%%%%%%%%%%%%%%%%%%%%%%%%%%%%%%%%%%%%%%%%
	%%%%%%%%%%%%%%%%%%%%%%%%%%%%%%%%%%%%%%%%%%%%%%%%%%%%%%%%%%%%%%%%%%%%%%%%%%%%%%%%%%%%%%%%%%%%
	
				%%%%%%%%%%%%%%%%%%%%%%%%%%%%%%%%%%%%%%%%%%%%%%%%%%%%%%%%%%%%%%%%%%%%%%%%%%%%%%%%%%%%%%%%%%%%%
	\begin{figure}[h]
		\begin{center}
			\includegraphics[width=8.7cm]{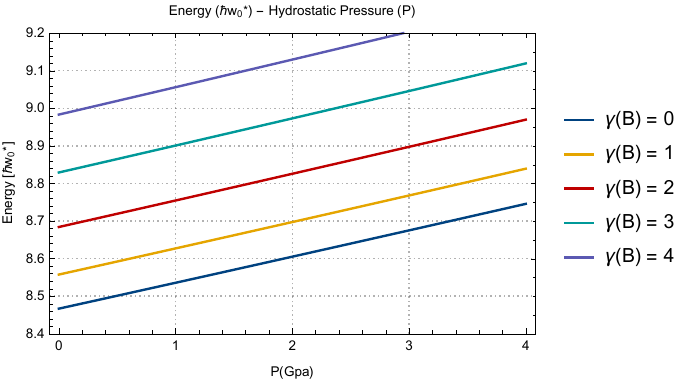}
		\end{center}
	\vspace{-0.5cm}
		\caption{Energy as a function of the pressure for differents magnetic field $\B$.}
		\label{2}
	\end{figure}
	%%%%%%%%%%%%%%%%%%%%%%%%%%%%%%%%%%%%%%%%%%%%%%%%%%%%%%%%%%%%%%%%%%%%%%%%%%%%%%%%%%%%%%%%%%%%
	
	As the magnetic field changes over the quantum dot, the energy levels intersect. As a result of the change in the orientation of the magnetic moments (parallel or antiparallel), the change of sign of $m$ produces a splitting of the energy levels,  which translates to an
increase or decrease of energy. This particular phenomenon was described analytically by Fock\cite{fock} and Darwin\cite{darwin}, and indicates the values of $m$ $(\pm)$ for which the system energy is different. In Figure \ref{1} we plot a single energy level of the electron at the quantum dot and its vertical upward displacement as the hydrostatic pressure increases. This phenomenon suggests that it is possible to use hydrostatic pressure as a tool to generate optical transitions between energy levels for the electron and, therefore use it as a control mechanism for the implementation of quantum gates of 1-qubit\cite{loss}.

				%%%%%%%%%%%%%%%%%%%%%%%%%%%%%%%%%%%%%%%%%%%%%%%%%%%%%%%%%%%%%%%%%%%%%%%%%%%%%%%%%%%%%%%%%%%%%
\begin{figure}[h]
	\begin{center}
		\includegraphics[width=8.7cm]{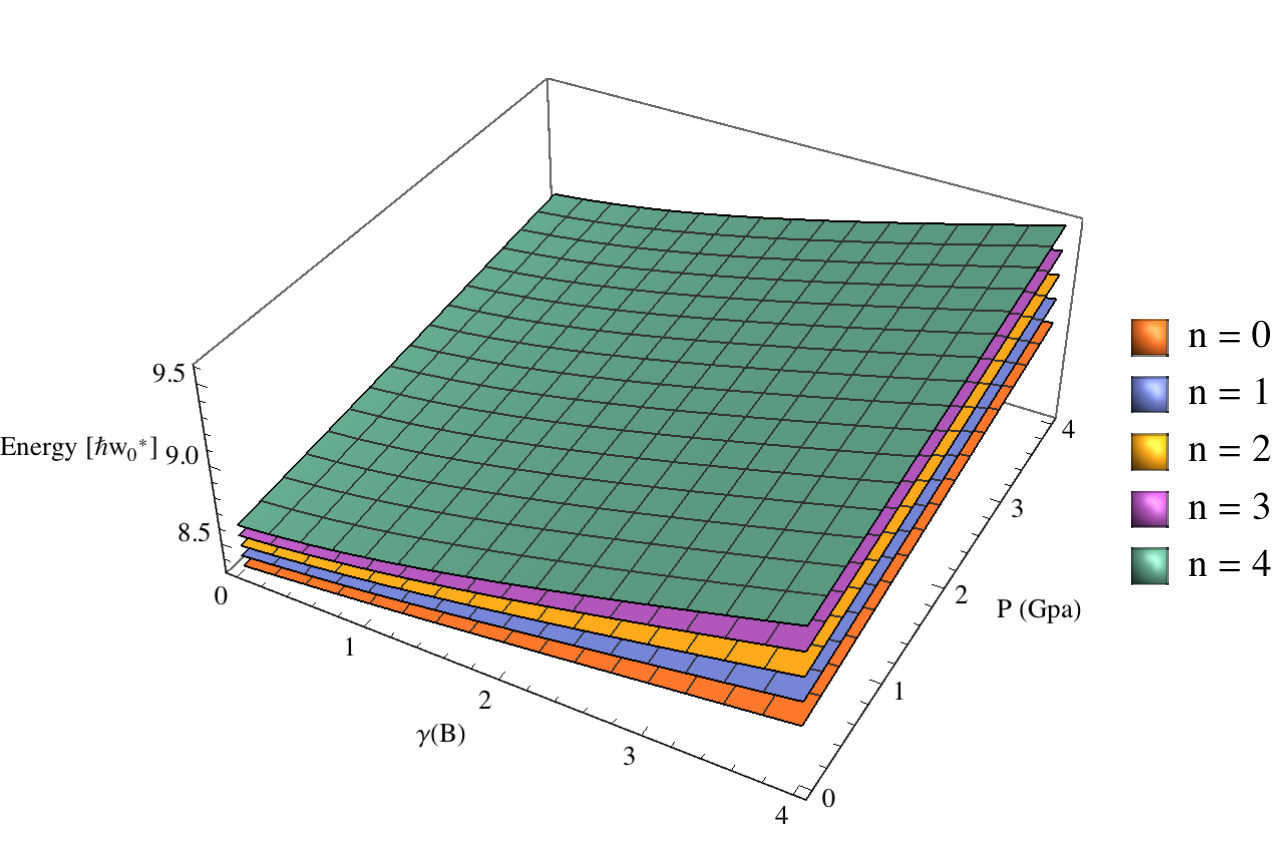}
	\end{center}
	\vspace{-0.5cm}
	\caption{Energy as a function of the pressure and $\gamma (\B)$ for differents $n$ levels.}
	\label{3}
\end{figure}
%%%%%%%%%%%%%%%%%%%%%%%%%%%%%%%%%%%%%%%%%%%%%%%%%%%%%%%%%%%%%%%%%%%%%%%%%%%%%%%%%%%%%%%%%%%%

We show the variation of the energy with pressure in Figure \ref{3}. It is clear that the energy increases linearly with the strenght of the hydrostatic pressure. Comparing the energy spectra of the
quantum dot under the effect of pressure (Figure \ref{2}) with the case of no external pressure ($P = 0$) given in Figure \ref{1}, we can see that there is a increase  in the energy spectrum. This behavior is explained with help of the dependence of the effective mass of the confined electron on the pressure given in equation \ref{eqb5}.

				%%%%%%%%%%%%%%%%%%%%%%%%%%%%%%%%%%%%%%%%%%%%%%%%%%%%%%%%%%%%%%%%%%%%%%%%%%%%%%%%%%%%%%%%%%%%%
\begin{figure}[h]
	\begin{center}
		\includegraphics[width=8.7cm]{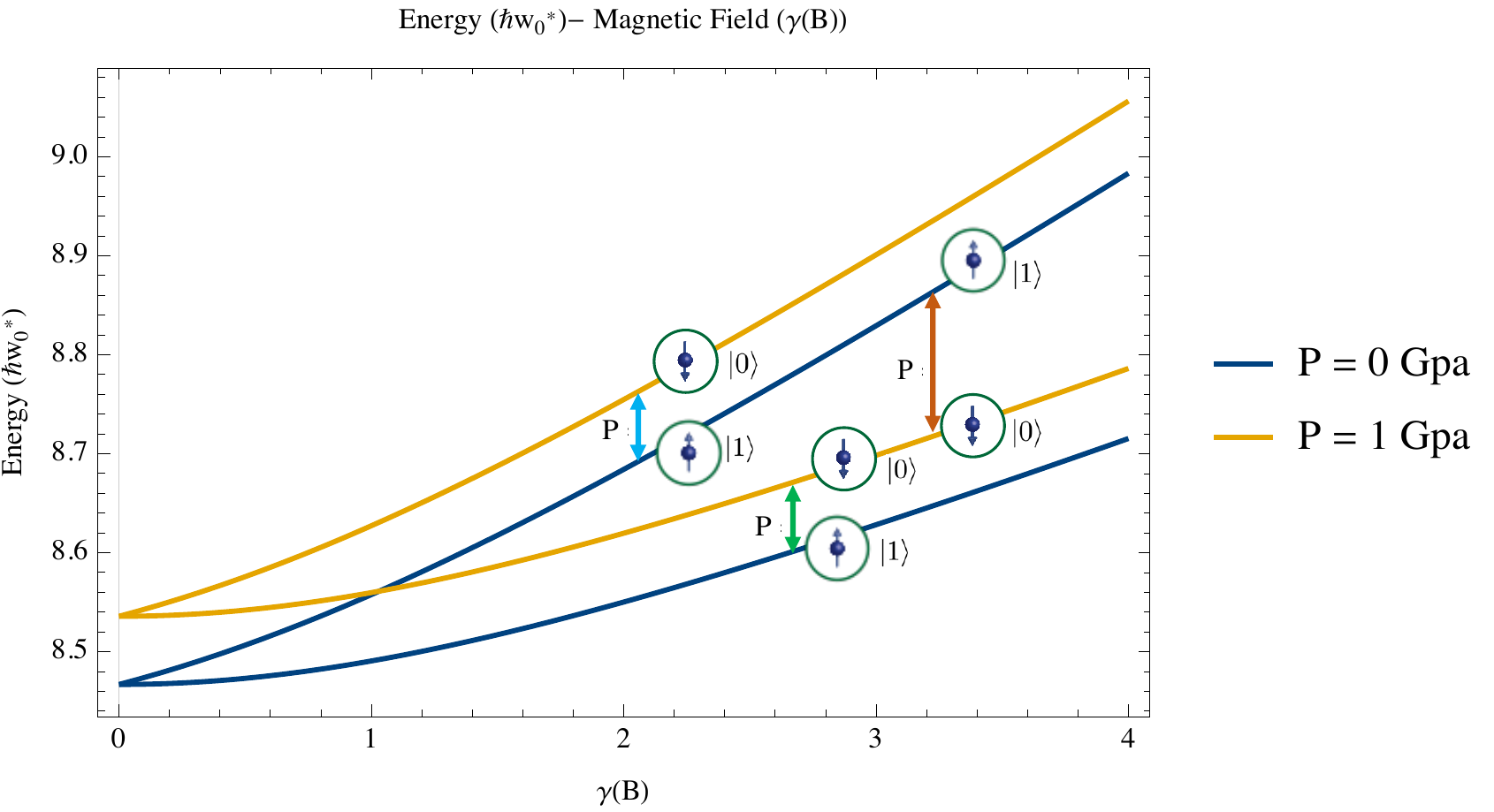}
	\end{center}
	\vspace{-0.5cm}
	\caption{Proposal of a quantum NOT gate using variations of hydrostatic pressure at a quantum dot.}
	\label{4}
\end{figure}
%%%%%%%%%%%%%%%%%%%%%%%%%%%%%%%%%%%%%%%%%%%%%%%%%%%%%%%%%%%%%%%%%%%%%%%%%%%%%%%%%%%%%%%%%%%%
 
Figure \ref{4} shows the schematic of a 1-qubit quantum gate. By changing the hydrostatic pressure, the electron confined at the quantum point changes energy level, operating as a NOT gate. 
	\section{Conclusions}
	\label{conclusiones}
	
Using the effective mass approximation and the algebraic formalism of creation and destruction operators, we determined the energy spectrum for a quantum dot of GaAs with an electron, under the effects of a constant external magnetic field $\B$ and  hydrostatic pressure.

The hydrostatic pressure modifies the single-electron energy spectrum. In general, the application of hydrostatic pressure in the quantum dot increases the energy. This result offers a new possibility to implement and complete a 1-qubit quantum gate controlled with variations of hydrostatic pressure. The confinement in a narrow dot system operating under hydrostatic pressure and a constant magnetic field may be used to tune the output of an optoelectronic device while modifying the physical size of quantum dots. Specifically, a scheme for the implementation of a negation gate was proposed, using variations in the hydrostatic pressure as a control system over the quantum dot.

\section{Acknowledgements}
\label{acknowledgements}

This work was partially supported by the Instituto Polit\'ecnico Nacional (IPN) of M\'exico, with the research project SIP-IPN 20171070 and by COFAA-IPN.

\bibliography{articulo-qd}

\end{document}